\documentclass[pra,twocolumn,showpacs,showkeys]{revtex4-1}

\usepackage{amssymb,amsfonts,amsmath,enumerate}
\usepackage{graphics}

\def\B{{\mathcal B}}
\def\F{{\mathcal F}}
\def\H{{\mathcal H}}

\def\E{{\mathcal E}}
\def\M{{\mathcal M}}
\def\K{{\mathcal K}}
\def\N{{\mathcal N}}
\def\G{{\mathcal G}}
\def\R{{\mathcal R}}
\begin{document}

\title{Comment  on ``Improved  bounds  on entropic  uncertainty relations''}

\author{G.M.~Bosyk, M. Portesi, A. Plastino}
\affiliation{Instituto  de F\'isica  La  Plata (IFLP,  CONICET),  and Dpto.\  de
  F\'isica, Fac.\ de Ciencias Exactas, Universidad Nacional de La Plata, 1900 La
  Plata, Argentina}
\author{S.~Zozor} \affiliation{Laboratoire Grenoblois d'Image, Parole, Signal et
  Automatique  (GIPSA-Lab, CNRS),  961 rue  de la  Houille Blanche,  38402 Saint
  Martin d'H\`eres, France}

\begin{abstract}
  We provide an analytical proof of the entropic uncertainty relations presented
  by de Vicente and  S\'anchez-Ruiz in [Phys.~Rev.~A \textbf{77}, 042110 (2008)]
  and also show  that the replacement of Eq.~(27) by  Eq.~(29) in that reference
  introduces solutions that do not take fully into account the contraints of the
  problem, which in turn leads to some mistakes in their treatment.
\end{abstract}

\maketitle

Consider two observables  $A$ and $B$, nondegenerate, with  discrete spectra and
complete orthonormal sets of  eigenvectors $ \left\{ \left\vert a_i\right\rangle
\right\}_{i=1}^N$ and  $\left\{ \left\vert b_j  \right\rangle \right\}_{j=1}^N$,
respectively.   Denote  by $  p_i(A)  =  \left\vert  \langle a_i  |\Psi  \rangle
\right\vert^2$  the  probabilities  for  the  outcomes of  observable  $A$  (and
analogously for $B$) when the system  is in the (pure) quantum state $\left\vert
  \Psi  \right\rangle $.   Let  $c  = \max_{i,j}  \left  \vert \left\langle  a_i
  \right\vert \left. b_j \right\rangle \right\vert \in \left[ \frac{1}{\sqrt{N}}
  ; 1  \right]$ be the  so called overlap  between the observables.  Maassen and
Uffink~(MU)~\cite{MaaUff88} proved  a non trivial universal lower  bound for the
sum
\begin{equation}
H(A) + H(B) \geq - 2 \ln c = \B_{MU},
\label{MUEUR:q}
\end{equation}
where  $H =  -  \sum_i p_i  \ln  p_i$ denotes  Shannon  entropy.  This  entropic
uncertainty  relation  (EUR)  has  also  been proved  by  Bialynicki-Birula  and
Mycielski~\cite{BiaMyc75}  in  the   special  case  when  the  observables  are
conjugated, namely $\langle a_i \vert \Psi \rangle $ and $\langle b_i \vert \Psi
\rangle $  are linked by  a Fourier  transform; then the  bound is sharp  in the
sense that there  exists a state $\left\vert \Psi \right\rangle  $ for which the
inequality is saturated, with $H(A) + H(B) = \ln N$.

In Ref.~\cite{VicSan08}, de Vicente and S\'anchez-Ruiz present an improvement of
the MU-EUR \eqref{MUEUR:q}, showing numerically  that $H(A) + H(B) \geq \B_{VS}$
where the bound reads
\begin{equation}
\B_{VS} = \left\{
\begin{array}{cll}
- 2 \ln c & \mbox{ if } & \displaystyle 0 < c \leq \frac{1}{\sqrt{2}}\vspace{.5mm}\\
\H_1(c) & \mbox{ if } & \displaystyle \frac{1}{\sqrt{2}} \leq c \leq c^\ast\vspace{1.5mm}\\
\F(c) & \mbox{ if } & c^\ast \leq c \leq 1
\end{array}
\right.
\label{VSREUR:eq}
\end{equation}
with
\begin{equation}
\F(c) = - (1+c) \ln \left( \frac{1+c}{2} \right) - (1-c) \ln \left(
\frac{1-c}{2}\right)
\label{F:eq}
\end{equation}
and
\begin{eqnarray}
\H_1(c) & = & - P_A \ln P_A - (1-P_A) \ln (1-P_A)  \notag \\
&& - P_B \ln P_B - (1-P_B) \ln (1-P_B)
\label{H1:eq}
\end{eqnarray}
where
\begin{equation}
P_A \equiv \cos^2 \alpha, \quad P_B \equiv \cos^2(\theta -\alpha ),
\quad c \equiv \cos \theta,
\label{changeVar:eq}
\end{equation}
and $\alpha $ is a (numerical) solution of the equation
\begin{eqnarray}
0 & = & \sin (2\alpha )\ln \left( \frac{1+\cos (2\alpha )}{1-\cos (2\alpha )}
\right) + \notag \\
& + & \sin (2(\alpha -\theta )) \ln \left( \frac{1+\cos (2\,(\alpha -\theta
))}{2\,\left( 1- \cos^2(\alpha -\theta )\right) }\right)
\label{EqSin:eq}
\end{eqnarray}
where  $\alpha \ne  \theta/2$ and  $\alpha  \ne \theta/2  + \pi/4$  in order  to
specify  $P_A  \ne  P_B$.   The  approximate value  of  $c^\ast$  is  determined
numerically in~\cite{VicSan08}.

We   show  that  the   replacement  of   Eq.~(27)  of   Ref.~\cite{VicSan08}  by
Eq.~\eqref{EqSin:eq},   via   the   change  of   variables~\eqref{changeVar:eq},
introduces solutions that do not take  fully into account the constraints of the
problem. This could potentially lead to erroneous conclusions.  In the sequel we
provide an analytical  proof of such results, discussing  in detail all possible
cases. At the end of the Comment we give some concluding remarks.

\hfill

The    mechanism    proposed    in    Ref.\cite{VicSan08}   to    improve    the
bound~\eqref{MUEUR:q} introduces the Landau--Pollak inequality~(LPI)
\begin{equation}
\arccos \sqrt{P_A} + \arccos \sqrt{P_B} \geq \arccos c,
\label{LP:eq}
\end{equation}
where $P_I = \max_i p_i(I)$ for $I =  A, B$, in two steps.  First, for $I=A$ and
$B$,  minimize Shannon  entropy $H(I)$  subject to  a fixed  maximum probability
$P_I$, which leads to the  minimal entropies $H_{\min}(P_I)$; second, search for
the infimum of
\begin{equation}
\M(P_A,P_B) = H_{\min}(P_A) + H_{\min}(P_B),
\label{SumMini:eq}
\end{equation}
over  the  possible  $P_A$  and   $P_B$,  subject  to  the  LPI.   According  to
Ref.~\cite{VicSan08},  the normalized  probability  distribution that  minimizes
$H(I)$ subject  to fixed  $P_I$ is  $( \underbrace{P_I ,  \ldots ,  P_I}_{M_I} ,
1-M_I P_I , 0 , \ldots , 0 ) $, where $M_I$ is a positive integer such that
\begin{equation}
\frac{1}{M_I+1} < P_I \leq \frac{1}{M_I}.
\label{ConstraintsM:eq}
\end{equation}
Then, one has
\begin{equation}
H_{\min}(P_I) = - M_I P_I \ln P_I - (1-M_I P_I) \ln (1-M_I P_I)
\label{Hmin:eq}
\end{equation}
for $I=A$ and $B$, and the  minimization of $\M$ is restricted by the inequality
constraints~\eqref{LP:eq}  and~\eqref{ConstraintsM:eq}.    We  present  rigorous
solutions to the problem, for different cases.

\hfill

\noindent {\bf Case $P_A\neq \frac{1}{M_A}$ and $P_B\neq \frac{1}{M_B}$}: \\
The      minimization     is      solved     introducing      the     Lagrangian
${\displaystyle{\sum_{I=A,B}}}   \left[  H_{\min}(I)  +   \mu_I  \left(   P_I  -
    \frac{1}{M_I} \right) + \nu_I \left( \frac{1}{M_I+1} - P_I \right) \right] +
\lambda \left( \arccos c - \arccos \sqrt{P_A} - \arccos \sqrt{P_B} \right)$ with
Lagrange parameters $\mu_A, \mu_B, \nu_A,  \nu_B$ and $\lambda $.  Deriving with
respect to  $P_A$ and $P_B$ and using  Karush--Khun--Tucker necessary conditions
for a minimum, one has
\begin{equation}
- M_I \ln \frac{P_I}{1-M_IP_I} + \frac{\lambda }{2 \sqrt{P_I \left( 1 - P_I
\right)}} + \mu_I - \nu_I = 0,
\label{DiffL:eq}
\end{equation}
\begin{equation}
\mu_I \left( P_I - \frac{1}{M_I} \right) = 0, \quad \nu_I \left( \frac{1}{M_I+1}
- P_I \right) = 0,
\label{ConstraintsMu:eq}
\end{equation}
\begin{equation}
\lambda \left( \arccos c - \arccos \sqrt{P_A} + \arccos \sqrt{P_B} \right) = 0
\label{ConstraintLP:eq}
\end{equation}
with $\mu_I, \nu_I, \lambda \geq 0$, for $I=A$ and $B$.

Since    the    inequalities~\eqref{ConstraintsM:eq}   are    strict,
from~\eqref{ConstraintsMu:eq}  one has $\mu_I  = \nu_I  = 0$.   It is  proved in
Ref.~\cite{VicSan08}  that  $\lambda  \neq  0$  (otherwise  $P_I  =  1/(M_I+1)$);
therefore, the LPI becomes an equality.   Taking the cosine of this equality and
using the constraints~\eqref{ConstraintsM:eq} one has
\begin{eqnarray}
c & = & \sqrt{P_AP_B}-\sqrt{\left( 1-P_A\right) \left( 1-P_B\right)}
\label{EqC:eq} \\
& \leq & \frac{1-\left( M_A-1\right) \left( M_B-1\right) }{\sqrt{M_A M_B}}.
\label{IneqC:eq}
\end{eqnarray}
As $M_I$ is  a positive integer and $c  > 0$, at least one $M_I$  must be unity,
and then
\begin{equation}
c \leq \frac{1}{\sqrt{\max \left( M_A , M_B\right)}}.
\label{min1Cmax:eq}
\end{equation}
One  can  assume   $M_A=1$  and  $M_B=M\geq  1$.  Extracting   $\lambda  $  from
Eqs.~\eqref{DiffL:eq} when $I=A$ and $B$, one gets
\begin{eqnarray}
& \sqrt{P_A\left( 1-P_A\right) } & \ln \left( \frac{P_A}{1-P_A} \right) =
\notag\\
M & \sqrt{P_B\left( 1-P_B\right) } & \ln \left( \frac{P_B}{1-MP_B} \right).
\label{DiffL_M:eq}
\end{eqnarray}
These equations  have \textit{several solutions}.   For example, a  solution for
$M=1$ is given by $P_A = P_B = \frac{1+c}{2}$, making the function $\F$ given in
Eq.~\eqref{F:eq} a possible candidate for a lower bound of the entropy sum.  For
$P_A  \neq P_B$,  Eqs.~\eqref{EqC:eq}--\eqref{DiffL_M:eq} do  not  have analytic
solutions. At this stage, the authors in Ref.~\cite{VicSan08} perform the change
of  variables~\eqref{changeVar:eq} and  solve  numerically Eq.~\eqref{EqSin:eq},
instead  of Eq.~\eqref{DiffL_M:eq}  with  $M=1$, proposing  the function  $\H_1$
given  in Eq.~\eqref{H1:eq}  as  a  possible minimum  in  a range  $0  < c  \leq
c^\ast$. This is the critical point that motivates this Comment.

We will  present below  a detailed analysis  that exhibits the  following facts,
depending on the value of the overlap:
\begin{enumerate}
\item In the range $0 < c < \frac{1}{\sqrt2}$:
\begin{enumerate}[\hspace{-7.5mm}$\bullet$]
\item For  $M = 1$ Eq.~\eqref{DiffL_M:eq}  has only the trivial  solution $P_A =
  P_B$, leading to  $\F$.  The value of $\H_1$  reported in Ref.~\cite{VicSan08}
  corresponds to  $P_A$ and $P_B$ outside  the allowed interval  [see Eqs.  eqs.
  \eqref{ConstraintsM:eq} and  \eqref{EqC:eq}].  In fact, Eq.~\eqref{DiffL_M:eq}
  for $M=1$ (together with \eqref{ConstraintsM:eq} and \eqref{EqC:eq}) have only
  the trivial solution $P_A = P_B$, leading to $\F$.  This seems to open the way
  to improve the MU bound in this range.
\item However,  the extremum  attained at $\frac{1+c}{2}$  happens to be  a {\it
    maximum} for $\M$. If this value lies below the MU bound, so is the minimum;
  but,  for the  range $c  \in  ( c^\dag  ; \frac{1}{\sqrt{2}}  )$ with  $c^\dag
  \approx 0.61$, the maximum is higher than $ -2\ln c$.
\item In  fact $P_A$ (resp.\  $P_B$) ``lives'' within  a given interval  and the
  minimum  of  $\M$ is  attained  {\it  at the  end  points}  of that  interval.
  Moreover,  this minimum is  less than  $- 2  \ln c  $, which  analytically and
  rigorously  proves  the   result  of Ref.~\cite{VicSan08}  in  $0   <  c  \leq
  \frac{1}{\sqrt{2}}$.
\end{enumerate}
\item  In   the  range  $\frac{1}{\sqrt{2}}   \leq  c  \leq   c^\ast$:  solution
  $\frac{1+c}{2}$ still  corresponds to  a maximum for  $\M(P_A,P_B)$.  However,
  Eq.~\eqref{DiffL_M:eq} (with  $M = 1$) admits {\it  two symmetrical solutions}
  yielding the same minimum $\H_1$.   We prove analytically that the extremizing
  values of $P_A$ and  $P_B$ satisfy the constraints \eqref{ConstraintsM:eq} and
  \eqref{EqC:eq}.   However,   the  value   $\H_1(c)$  can  be   evaluated  only
  numerically.  The  result given in~\cite{VicSan08}  is then confirmed  in this
  range.  In passing, we prove that $c^\ast$ is a solution of the transcendental
  equation
\begin{equation}
c^\ast \ln \left( \frac{1+c^\ast}{1-c^\ast}\right) = 2 \quad \Leftrightarrow
\quad c^\ast \, \arg \!\tanh c^\ast = 1, \hspace{-6mm}
\label{cstar:eq}
\end{equation}
then $c^\ast \approx 0.834$ as found in Ref.~\cite{VicSan08}.
\item  In the  range $c^\ast  <  c \leq  1$: only  the solution  $\frac{1+c}{2}$
  remains, as  observed in Ref.~\cite{VicSan08}. Moreover,  it corresponds there
  to a minimum.  We justify this {\it analytically}, confirming
  the result of Ref.~\cite{VicSan08}
\end{enumerate}

\hfill

The  proofs  are  as follows.  First,  we  rewrite  Eq.~\eqref{EqC:eq} as  $c  +
\sqrt{(1-P_A)(1-P_B)} = \sqrt{P_A P_B}$. As both sides are positive, they can be
squared without  further ado leading  to a quadratic equation  in $\sqrt{1-P_B}$
whose only allowed solution is
\begin{equation}
\sqrt{1-P_B} = \sqrt{P_A(1-c^2)}-c\sqrt{1-P_A}.
\label{UnMoinsPb:eq}
\end{equation}
Solving for $P_B$ for given $c$ gives
\begin{equation}
P_B (P_A) = \left( \sqrt{\left( 1 - c^2 \right) \left( 1 - P_A \right) } + c
\sqrt{P_A} \right)^2.
\label{Pb:eq}
\end{equation}
We  realize  that  $P_A$,  appart  from  lying between  $\frac12$  and  1  (from
\eqref{ConstraintsM:eq} since  $M_A=1$), is constrained to be  larger than $c^2$
(from  the   positivity  of  \eqref{UnMoinsPb:eq}).    Furthermore,  the  bounds
\eqref{ConstraintsM:eq}   for   $I=B$   applied  to~\eqref{UnMoinsPb:eq}   yield
additional constraints on $P_A$. Summarizing, when $M=1$, we get
\begin{equation}
\left( P_A^- ; P_A^+ \right) = \left\{
\begin{array}{cll}
\displaystyle \left( \frac12 ; \frac{\left( c + \sqrt{1-c^2} \right)^2}{2}
\right) & \mbox{ if } & \displaystyle 0 < c < \frac{1}{\sqrt{2}} \\
\displaystyle \left( c^2 ; 1 \right) & \mbox{ if } & \displaystyle
\frac{1}{\sqrt{2}} \leq c\leq 1.
\end{array}
\right.
\label{ConstraintPa:eq}
\end{equation}
Next,  we consider  $\M_M(P_A) =  \M(P_A,P_B(P_A))$. The  goal is  to  study its
behavior versus  $P_A$ so as to determine  its minimum.  To such  end we compute
successive derivatives  of $\M_M$,  with the help  of some  auxiliary functions.
Noting      that,       for      given      $c$,       $\frac{dP_B}{dP_A}      =
-\frac{\sqrt{P_B(1-P_B)}}{\sqrt{P_A(1-P_A)}}$   we   obtain   $\M_M'   (P_A)   =
\frac{\E_M(P_A)}{\sqrt{P_A(1-P_A)}}$, where
\begin{eqnarray}
\E_M(P_A) & = & M \sqrt{P_B(1-P_B)} \, \ln \left( \frac{P_B}{1 - M P_B} \right)
\notag \\
&& - \sqrt{P_A(1-P_A)} \, \ln \left( \frac{P_A}{1 - P_A} \right).
\label{FunctionE:eq}
\end{eqnarray}
has  the  same  sign  as  $\M_M'$.   It  is obvious  that  setting  $\E_M  =  0$
solves~\eqref{DiffL_M:eq}.  In the sequel, let us restrict ourselves to the case
$M  =  1$ (we  will  confirm  later that  the  cases  $M >  1$  need  not to  be
considered). We  demonstrate now  that $\E_1$ has  only four types  of behavior.
Its derivative writes as $\E_1'(P_A) =  - \, \frac{\K(P_A)}{2\sqrt{P_A ( 1 - P_A
    )}}$, where
\begin{eqnarray}
\K(P_A) & = & \left( 1 - 2 P_B \right) \, \ln \left( \frac{P_B}{1 - P_B} \right)
\notag \\
& + & \left( 1 - 2 P_A \right) \, \ln \left( \frac{P_A}{1-P_A} \right) + 4,
\label{FunctionK:eq}
\end{eqnarray}
has the  opposite sign of $\E_1'$.   We will see  that $\K$ has always  the same
behavior versus  $P_A$ independently of  $c$: it increases  up to a  maximum and
then decreases.  This behavior, together with  the sign of the  maximum of $\K$,
completely determines the shape of $\E_1$.  The derivative of $\K$ is $\K' (P_A)
= \frac{\N(P_A)}{\sqrt{P_A(1-P_A)}}$, where
\begin{equation}
\begin{array}{lll}
\N(P_A) & = & \displaystyle 2 \sqrt{P_B ( 1 - P_B )} \ln \left(
\frac{P_B}{1-P_B} \right) - \frac{1 - 2 P_B}{\sqrt{P_B ( 1 - P_B)}}
\\
& - & \displaystyle 2 \sqrt{P_A ( 1 - P_A )} \ln \left( \frac{P_A}{1 - P_A}
\right) + \frac{1 - 2 P_A}{\sqrt{P_A ( 1 - P_A )}}
\end{array}
\nonumber
\end{equation}
has  the same sign  as $\K'$.  Finally, the  derivative of  $\N$ is  $\N'(P_A) =
\frac{dP_B}{dP_A} \frac{ (  1 - 2 P_B )  \R(P_B) + 4}{\sqrt{P_B ( 1 -  P_B )}} -
\frac{( 1 -  2 P_A ) \R(P_A) + 4}{\sqrt{P_A  ( 1 - P_A )}}$,  where $\R(x) = \ln
\left( \frac{x}{1-x} \right) + \frac{1 - 2 x}{2 x (1 - x)}$. From the negativity
of $\R(x)$  for $x \in  (\frac12 ; 1)$  and of $\frac{dP_B}{dP_A}$,  we conclude
that $\N' < 0$. Summing up,  for any $c$, $\N$ is continuous, strictly decreases
in $( P_A^- ; P_A^+)$, and  has only one root: $\N\left( \frac{1+c}{2} \right) =
0$.  As  a  result, $\K$  increases  with  $P_A$  in  the  interval $(  P_A^-  ;
\frac{1+c}{2} )$ and decreases in  $( \frac{1+c}{2} ; P_A^+ )$.  Furthermore, we
notice that ${\displaystyle \lim_{P_A \to  P_A^-} \K(P_A) = \lim_{P_A \to P_A^+}
  \K(P_A)}$ due to the fact that when $P_A \to P_A^-$ then $ P_B \to P_A^+$, and
vice versa. With respect to the sign of $\K$, we show that only three situations
arise:
\begin{enumerate}
\item when $c \in \left( 0 ; \frac{1}{\sqrt{2}} \right) $,
  the maximum  of $\K$,  given by $\K\left(  \frac{1+c}{2} \right)  = - 2  c \ln
  \left( \frac{1+c}{1-c} \right) + 4$, is positive.
  Besides,  the  value of  $\K$  at the  end  points,  given by  ${\displaystyle
    \lim_{P_A \to P_A^\pm}} \K(P_A) = - 2 c \sqrt{1-c^2} \ln \left( \frac{ 1 + 2
      c \sqrt{1-c^2} }{ 1 - 2 c  \sqrt{1-c^2} } \right) + 4$, decreases with $c$
  from $4$ to $-\infty $ and thus can have either sign.
\item when $c \in \left( \frac{1}{\sqrt{2}} ; c^\ast \right) $,
  the  maximum of  $\K$ is  also  positive, while  $\displaystyle \lim_{P_A  \to
    P_A^\pm}  \K(P_A) =  -  \infty$.  We  can  determine the  value of  $c^\ast$
  precisely    when   the    maximum    $\K$   becomes    zero,   arriving    at
  Eq.~\eqref{cstar:eq}.
\item when $c \in  \left( c^\ast ; 1 \right) $, the  maximum of $\K$ is negative
  and thus $\K < 0$ for all $P_A$.
\end{enumerate}

Going back to functions $\E_1$ and $\M_1$ we conclude that:
\begin{enumerate}
\item when $c\in \left( 0 ; \frac{1}{\sqrt{2}} \right)$, if $\K(P_A^\pm) \ge 0$,
  $\E_1' \le  0$ and  thus $\E_1$  is strictly decreasing;  on the  contrary, if
  $\K(P_A^\pm) < 0$, $\E_1$ increases,  decreases, and again increases.  In both
  cases, $\displaystyle \lim_{P_A \to P_A^-}  \E_1(P_A) = - \lim_{P_A \to P_A^+}
  \E_1(P_A)   =  \frac{1   -   2  c^2}{4}   \ln   \left(  \frac{\sqrt{1-c^2}   +
      c}{\sqrt{1-c^2} - c}\right) > 0$; hence  $ \E_1= 0$ has only one solution,
  given by  $P_A = \frac{1+c}{2}$.  This justifies that  the value of  $\H_1$ is
  computed  from values  of  $P_A$ and  $P_B$  that do  not satisfy  constraints
  eq.~\eqref{ConstraintsM:eq}  in  this range.   Moreover,  $\E_1$ changes  from
  positive to negative sign at  $\frac{1+c}{2}$, implying that: (i) the extremum
  of $\M_1$ at $\frac{1+c}{2}$ corresponds in fact to a maximum as we previously
  claimed, and  (ii) the minimum of $\M_1$  would be attained at  the end points
  $P_A^\pm$.  As  a  conclusion,   $\M_1$  is  lower-bounded  by  $\displaystyle
  \lim_{P_A \to P_A^{\pm}} \M_1(P_A)$ given by
\begin{equation}
\begin{array}{lll}
\displaystyle \M_{\inf} & = & \displaystyle - \frac{1 + 2 c \sqrt{1-c^2}}{2} \ln
\left( \frac{1 + 2 c \sqrt{1-c^2}}{4}\right) \\
& & \displaystyle - \frac{1 - 2 c \sqrt{1-c^2}}{2} \ln \left( \frac{1 - 2 c
\sqrt{1-c^2}}{4}\right).
\end{array}
\label{M_edges:eq}
\end{equation}
We now define the difference of MU-bound to this infimum: $\Delta_{\M_{\inf}}(c)
=  \B_{MU}(c)  - \M_{\inf}(c)$.   The  derivative  of $\Delta_{\M_{\inf}}$  with
respect to $c$ is $\frac{1 -  2 c^2}{\sqrt{1-c^2}} \left[ \ln \left( \frac{1 + 2
      c \sqrt{1-c^2}}{1  - 2 c  \sqrt{1-c^2}} \right) -  \frac{2 \sqrt{1-c^2}}{c
    \left( 1 - 2 c^2 \right)}\right]$ which can be analytically proved easily to
be always negative in the range $c \in ( 0 ; \frac{1}{\sqrt 2} )$.  This implies
that $\Delta_{\M_{\inf}}(c)$ is decreasing,  with the lowest difference given by
$\displaystyle \lim_{c \to \frac{1}{\sqrt  2}^-} \Delta_{\M_{\inf}}(c) = \ln 2 >
0$. This  analytically proves  that $\M_{\inf} <  \B_{MU}$: it is  impossible to
improve the MU-EUR in the range $0 < c < \frac{1}{\sqrt{2}}$.  This confirms the
result of Ref.~\cite{VicSan08}.   Notice that studying what happens  for $M > 1$
or for $P_I = \frac{1}{M_I}$ is then not necessary.
\item \label{punto2} when $c\in \left(  \frac{1}{\sqrt{2}} ; c^\ast \right) $, $
  \E_1$ increases, decreases, and again increases, with $\displaystyle \lim_{P_A
    \to P_A^-} \E_1(P_A)  = - \lim_{P_A \to P_A^+} \E_1(P_A)  = - c \sqrt{1-c^2}
  \ln \left(  \frac{c^2}{1-c^2} \right)  < 0$;  hence $\E_1 =  0$ has  now three
  solutions:  $\frac{1+c}{2}$,  corresponding to  a  maximum  of $\M_1$  ($\E_1$
  locally  decreases), and  other two  giving the  same minimum  for  $\M_1$ (by
  symmetry).  The minimum value of $\M_1$  in this range is denoted as $\H_1$ in
  Ref.~\cite{VicSan08},  where  it is  obtained  after  solving numerically  for
  $\alpha$ in  Eqs.~\eqref{changeVar:eq}--\eqref{EqSin:eq}.  The same  result is
  obtained here directly from  \eqref{DiffL_M:eq} and \eqref{Pb:eq}, taking care
  of the constraints \eqref{ConstraintPa:eq} for $P_A$ (notice that only $M = 1$
  has   to    be   considered   as   \eqref{min1Cmax:eq}    enforces   $M   \leq
  \frac{1}{c^2}$). We also numerically confirm that  $\H_1(c) > - 2 \ln c$, thus
  giving the possibility of improving MU-EUR in this range (see the cases $P_I =
  1/M_I$ below).
\item when  $c \in  \left( c^\ast ;  1 \right)  $, $\E_1$ always  increases, the
  limiting values at  the end points are the same as  in \ref{punto2} above, but
  the  unique  root  $\frac{1+c}{2}$  corresponds  to a  minimum:  $\M_1  \left(
    \frac{1+c}{2} \right)  = \F(c)$ given in Eq.~\eqref{F:eq}  (notice that only
  $M = 1$ has to be  considered since $M \leq \frac{1}{c^2}$).  Consider now the
  difference   $\Delta_\F(c)   =  \B_{MU}(c)   -\F(c)$,   whose  derivative   is
  $\frac{2}{c} \left( c  \ln \left( \frac{1+c}{1-c} \right) - 2  \right) > 0$ in
  the range $c  > c^\ast$. Thus $\Delta_\F(c)$ increases;  since $\Delta_\F(1) =
  0$, then $\Delta_\F(c)<  0$ in this range and  therefore we analytically prove
  that  $\F > \B_{MU}$,  as observed  in \cite{VicSan08}.  Again the  MU-EUR can
  possibly be improved in this range (see the cases $P_I = 1/M_I$ below).
\end{enumerate}

What happens  in the particular cases  $c = \frac{1}{\sqrt2}$ and  $c = c^\ast$,
follows from the continuity of the functions involved.

\hfill

\noindent {\bf Cases $P_A = \frac{1}{M_A}$ and/or $P_B = \frac{1}{M_B}$}: \\
In Ref.~\cite{VicSan08}  the authors  check the case  $P_I =  \frac{1}{M_I}$ for
$I=A$ or $B$ and find as a possible minimum for $\M(P_A,P_B)$
\begin{equation}
\G(c) = - c^2 \left\lfloor \frac{1}{c^2} \right\rfloor \ln c^2 - \left(  1 - c^2
\left\lfloor \frac{1}{c^2} \right\rfloor \right)  \ln \left(1 - c^2 \left\lfloor
\frac{1}{c^2} \right\rfloor \right),
\label{G:eq}
\end{equation}
where $\lfloor \cdot  \rfloor $ indicates integer part  (floor). They base their
procedure on the equivalent of  Eqs.~\eqref{DiffL:eq} and claim that $M$ must be
either 1 or 2, assuming the nonnegativity of the Lagrange multipliers. As far as
we  understand, this  reasoning seems  erroneous, i.e.\  the  multiplier $\mu_I$
corresponding  to  the ``equality  constraint''  $P_I  =  \frac{1}{M_I}$ is  not
necessarily nonnegative, as should be for ``inequality constraints''.

First of all, as already mentioned, the  MU-EUR could be improved only when $c >
\frac{1}{\sqrt{2}}$.  For $P_A = \frac{1}{M_A}$ and $P_B \neq \frac{1}{M_B}$ (or
the contrary,  exchanging the  roles of  observables $A$ and  $B$), we  have for
$I=B$   the  same   Eqs.~\eqref{DiffL:eq}  and   \eqref{ConstraintsMu:eq};  also
Eq.~\eqref{ConstraintLP:eq} remains  valid. One still  has $\mu_B = \nu_B  = 0$,
and  thus $  \lambda  \neq 0$.   The  LPI remains  saturated,  which means  that
Eqs.~\eqref{EqC:eq}, \eqref{min1Cmax:eq}, \eqref{UnMoinsPb:eq} and \eqref{Pb:eq}
are   still    valid.    Thus,   the    constraint   $P_A   >   c^2$    due   to
Eq.~\eqref{UnMoinsPb:eq} enforces $P_A = 1$,  thus $M_A = 1$.  As a consequence,
Eq.~\eqref{Pb:eq}    gives   $P_B    =    c^2$,   which    means   also    that,
from~\eqref{ConstraintsM:eq}, $M_B  = \lfloor \frac{1}{c^2}  \rfloor$.  In turn,
one  arrives   at  the  function  $\G$   given  above,  valid  only   for  $c  >
\frac{1}{\sqrt{2}}$.   This  also entails  that  $  \left \lfloor  \frac{1}{c^2}
\right\rfloor = 1$ and then $\G = -  c^2 \, \ln c^2 - (1-c^2) \ln (1-c^2)$ which
corresponds,  as  it   should,  to  the  Shannon  entropy   of  the  probability
distribution $(c^2, 1-c^2,  0,\ldots, 0)$. One can numerically  prove that $\G >
\H_1$ and,  analytically, that for  $c \geq c^\ast$,  $\G > \F$;  therefore $\G$
does not correspond to the minimal $\M$.

Finally, we study the case $P_A  = \frac{1}{M_A}$ and $P_B = \frac{1}{M_B}$. The
sum of  the minimum Shannon entropies is  $\M(\frac{1}{M_A},\frac{1}{M_B}) = \ln
(M_A M_B)$. It is  straightforward to see that the minimal $\M$  is: $0$ if $c =
1$ or $\ln  M$ if $\frac{1}{\sqrt{M}} \leq c <  \frac{1}{\sqrt{M-1}}$ where $M =
2, 3, 4, \ldots$. Clearly these bounds are non-optimal.

\hfill

Summing up, we revisit analytically the full resolution of the problem presented
in  Ref.~\cite{VicSan08}  that  deals   with  the  uncertainty  related  to  the
measurement of  two discrete  quantum observables, using  as measure the  sum of
Shannon  entropies   associated  to   both  distributions  constrained   by  the
Landau-Pollak inequality.  De Vicente and S\'anchez-Ruiz show in~\cite{VicSan08}
that the  Maassen-Uffink bound  can be improved  using this constraint  when the
overlap $c$ between observables is in the range $( 1/\sqrt{2} ; 1 )$; we confirm
analytically  this  result.  Our   central  contributions  were  to  provide  an
analytical proof of the non-improvement of the bound when $c$ is in the range $(
0 , 1/\sqrt{2})$, and the analytical proof  that $\F$ is indeed a minimum of the
entropy sum $\M$ for  $c$ in the range $( c^* ;  1 )$. Additionally, we obtained
the value of $c^*$ from an analytical expression, given in Eq. \eqref{cstar:eq}.
We  detected a  mistake in  the  VS-treatment of  the constrained  extremization
problem:   the  function   $\mathcal{H}_1$   was  computed   for  solutions   of
Eq.~\eqref{EqSin:eq} that do not take into account the whole set of restrictions
on the pertinent probabilities. This seemed to open the possibility of improving
Maassen--Uffink bound  in the range $c\in  ( 0 ; \frac{1}{\sqrt{2}}  )$.  But in
fact, we rigorously show that it is impossible to improve MU-EUR with the LPI in
this range.

Moreover, let us comment that the function ${\mathcal{F}}(c)$ can be interpreted
as  half the  Jensen--Shannon  divergence  between the  pure  states $\vert  a_i
\rangle \langle a_i \vert$ and $ \vert b_j \rangle \langle b_j \vert$, for which
the  overlap is  maximum~\cite{Lam09}.   An interesting  future  research is  to
exploit this relationship for establishing new entropic uncertainty relations.

\hfill

The authors  acknowledge financial support  from CONICET and  ANPCyT (Argentina)
and CNRS (France).

\end{document}